\documentclass[journal=aesccq,manuscript=article,layout=twocolumn]{achemso}

\usepackage{graphicx}
\usepackage[table-align-text-post=false, range-units=single, range-phrase=\textendash, separate-uncertainty=false, multi-part-units=single, per-mode=symbol-or-fraction]{siunitx}
\usepackage{threeparttable}
\usepackage{booktabs}
\usepackage[section]{placeins}

\usepackage{hyperref}

\usepackage{lineno}
\DeclareSymbolFont{matha}{OML}{txmi}{m}{it}%
\DeclareMathSymbol{\varv}{\mathord}{matha}{118}

\usepackage{xcolor}
\definecolor{blue}{HTML}{648FFF}
\definecolor{pink}{HTML}{DC267F}
\definecolor{yellow}{HTML}{FFB000}

\usepackage[version=4]{mhchem}
\newcommand{\refsuppinfo}{Supporting Information}

\newcommand{\tnotes}[1]{{\vspace{0.02cm}\par\leftskip0.15cm \rightskip\leftskip\scriptsize\linespread{0.5}#1\newline\vspace{-\baselineskip}\par}}

\newcommand{\farcs}{.\!\!^{\prime\prime}}
\newcommand{\fs}{.\!\!^{\rm s}}

\author{Luis Bonah}
\email{bonah@ph1.uni-koeln.de}
\affiliation[Univsersity Cologne]
{I. Physikalisches Institut, Universität zu Köln, Zülpicher Str. 77, 50937 Köln, Germany}
\author{Jean-Claude Guillemin}
\affiliation[University Rennes]{Univ Rennes, Ecole Nationale Supérieure de Chimie de Rennes, CNRS, ISCR – UMR6226, 35000 Rennes, France}
\author{Arnaud Belloche}
\affiliation[MPI Bonn]{Max-Planck-Institut für Radioastronomie, Auf dem Hügel 69, 53121 Bonn, Germany}
\author{Sven Thorwirth}
\email{sthorwirth@ph1.uni-koeln.de}
\author{Holger S. P. Müller}
\author{Stephan Schlemmer}
\affiliation[Univsersity Cologne]
{I. Physikalisches Institut, Universität zu Köln, Zülpicher Str. 77, 50937 Köln, Germany}

\title[DR Spectroscopy of Glycidaldehyde]
  {Leveraging MMW-MMW Double Resonance Spectroscopy to Understand the Pure Rotational Spectrum of Glycidaldehyde and 17 of its Vibrationally Excited States}


\begin{document}

\begin{tocentry}
\includegraphics[width=\textwidth]{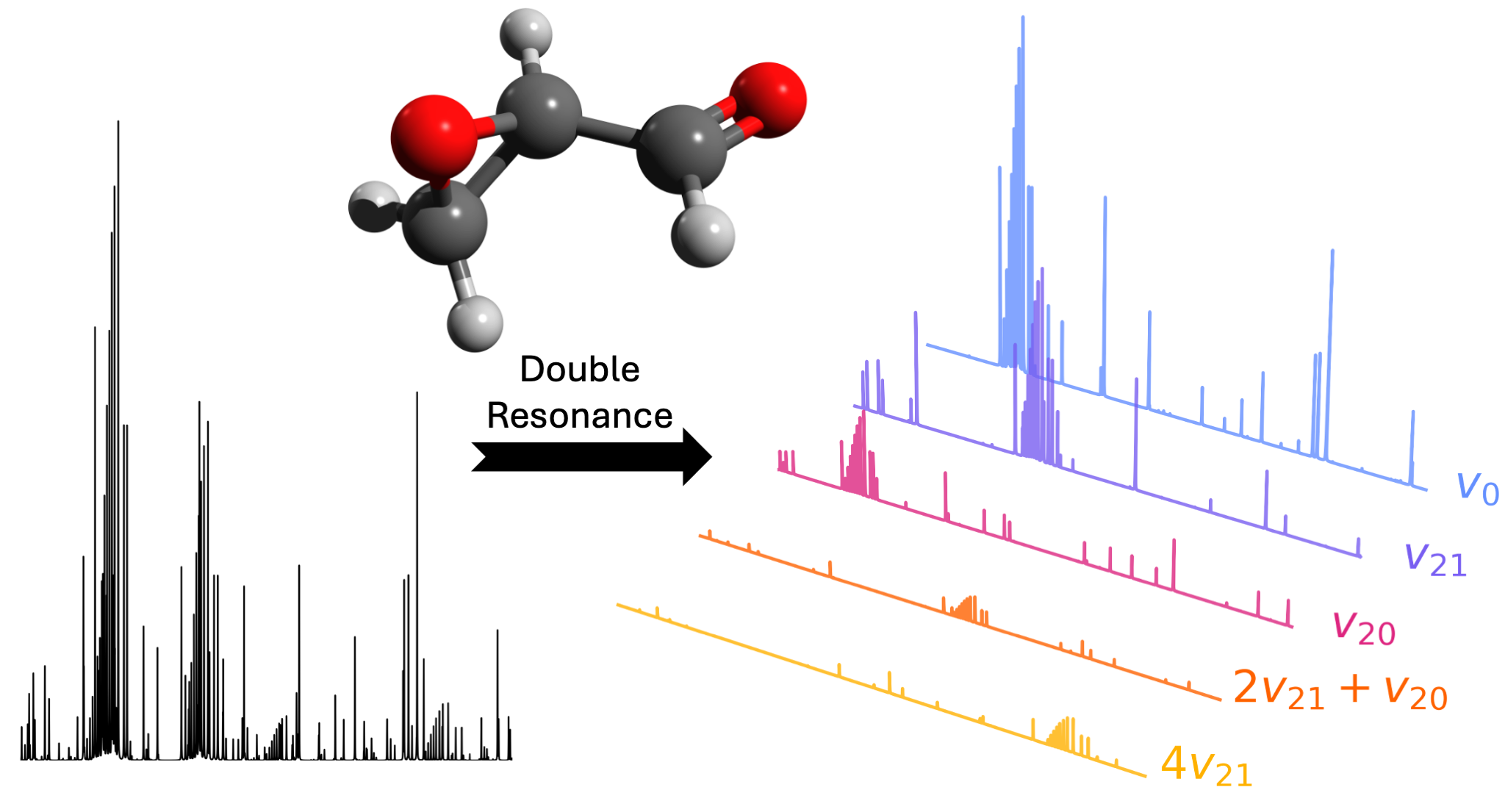}
\end{tocentry}

\begin{abstract}
Broadband measurements of glycidaldehyde in the frequency ranges \SIrange{75}{170}{GHz} and \SIrange{500}{750}{GHz} were recorded to extend previous analyses of its pure rotational spectrum in the microwave region.
The rotational parameters of the ground vibrational states for the main isotopologue and the three singly \ce{^{13}C} substituted isotopologues were considerably improved and additional higher-order parameters were determined.

To identify new vibrationally excited states in the dense and convoluted spectrum, an updated version of the double-modulation double-resonance spectroscopy technique was used.
Connecting transitions with a shared energy level into series and expanding these via Loomis-Wood plots proved to be a powerful method, which allowed the identification of 11 new vibrationally excited states in addition to the already known aldehyde torsions, $v_{21}=1$ to $v_{21}=6$.
Interactions between several vibrational states were observed and three interacting systems were treated successfully.

Rotational transitions of glycidaldehyde were searched for in the imaging spectral line survey ReMoCA obtained with the Atacama Large Millimeter/submillimeter Array (ALMA) toward the high-mass star-forming region Sgr~B2(N).
The observed spectra were modeled under the assumption of local thermodynamic equilibrium (LTE).
Glycidaldehyde, an oxirane derivative, was not detected toward Sgr~B2(N2b).
The upper limit on its column density implies that it is at least six times 
less abundant than oxirane in this source.

\end{abstract}

\section{Introduction}
\label{sec:Introduction}
To date, only a handful of ring molecules with heteroatoms in the ring have been detected in space.
Beginning with the smallest possible ring molecules, the three triatomics \ce{\textit{c}-SiC2}~\cite{Thaddeus1984}, \ce{\textit{c}-MgC2}~\cite{Changala2022}, and \ce{\textit{c}-CaC2}~\cite{Gupta2024} were first detected toward the carbon-rich asymptotic giant branch star CW Leo (IRC+10216) in 1984, 2022 and 2024, respectively.
\ce{SiC2} was found recently also toward the Galactic Center molecular cloud G+0.693–0.027~\cite{Massalkhi2023}.
Similarly, the four-atom molecule \ce{\textit{c}-SiC3} was detected in 1999 in the circumstellar envelope of CW Leo.
The two somewhat more complex ring molecules oxirane (\ce{\textit{c}-C2H4O}) and propylene oxide (\ce{CH3CHCH2O}) were first detected toward the high-mass star-forming region Sagittarius~B2(N) in 1997~\cite{Dickens1997} and 2016~\cite{McGuire2016}, respectively.
Oxirane was later also observed toward hot core regions~\cite{Nummelin1998}, Galactic center molecular clouds~\cite{RequenaTorres2008}, a low-mass protostar~\cite{Lykke2016}, and pre-stellar cores~\cite{Bacmann2019}.

The presence of oxirane in the interstellar medium gives rise to the question whether or not oxirane derivatives are astronomically abundant.
Glycidaldehyde (\ce{(C2H3O)CHO}), also known as oxiranecarboxaldehyde or 2,3-epoxy-propanal, is such a derivative in which one of oxirane's \ce{H} atoms is substituted by a \ce{-CHO} functional group.
Its rotational spectrum is known from a previous study in the microwave range (\SIrange{8}{40}{GHz}) by Creswell et al.\ from 1977~\cite{Creswell1977}.
There, the authors analyzed the ground vibrational states of the main isotopologue and the three singly substituted \ce{^{13}C} isotopologues in natural abundance.
Additionally, the lowest vibrational fundamental $\nu_{21}$, the aldehyde torsion at about \SI{125}{cm^{-1}}, was followed up to $v_{21}=7$.
However, even for the ground vibrational state only quartic centrifugal distortion parameters were determined, making frequency extrapolations to higher frequency ranges rapidly inaccurate.

In the present study, the frequency coverage was extended to the millimeter and sub-millimeter ranges reaching frequencies as high as \SI{750}{GHz}.
The dense and convoluted spectrum of glycidaldehyde makes it an ideal case study for double-modulation double-resonance (DM-DR) spectroscopy~\cite{Zingsheim2021}, which allows to filter the spectrum for lines sharing an energy level.
This simplified the analysis immensely, proved critical in identifying new vibrationally excited states, and understanding interactions between them.
In total, 17 vibrationally excited states of the main isotopologue were analyzed along with the ground vibrational states of the main isotopologue and the three singly substituted \ce{^{13}C} isotopologues.

\section{Experimental Details}
\label{sec:Experimental Details}
Broadband as well as double-resonance (DR) measurements were recorded at two experimental setups in Cologne with a sample synthesized on a gram scale according to a previously reported procedure~\cite{Payne1958}.
The broadband measurements cover the frequency ranges \SIrange{75}{170}{GHz} and \SIrange{500}{750}{GHz}.
Double-resonance (DR) measurements in the frequency range \SIrange{75}{120}{GHz} were used to confirm relationships between lines, identify new series of transitions, and find pure rotational transitions between vibrationally excited states (so-called interstate transitions) that arise due to wavefunction mixing when vibrational states interact with each other. 
The DR measurements were performed with a modified version of the double-modulation double-resonance (DM-DR) method (see \autoref{sec:DMDRScheme}) described previously~\cite{Zingsheim2021}.

\subsection{Experimental Setups}
\label{sec:Experiments}
Two absorption experiments were used to record the broadband spectra, each consisting of a source, an absorption cell, and a detector.
The sources consist of synthesizers with subsequent amplifier-multiplier chains.
The radiation is guided through the absorption cells and into the detector via lenses, mirrors, and horn antennas.
For the higher frequency range of \SIrange{500}{750}{GHz}, the absorption cell consists of a single \SI{5}{m} borosilicate glass cell in a double-pass setup for a total absorption path of \SI{10}{m}.
The lower frequency range setup (\SIrange{75}{170}{GHz}) uses two \SI{7}{m} borosilicate glass cells in a single-pass configuration for a total absorption path of \SI{14}{m}.
Double-pass configurations are foregone here as this lower-frequency experiment is also used for the DR measurements where the second polarization direction is used to align the pump and probe sources radiation co-spatially (see \autoref{sec:DMDRScheme}).
Different Schottky detectors were used for frequencies below \SI{500}{GHz} and a cryogenically cooled hot-electron bolometer (QMC QNbB/PTC(2+XBI)) was used for measurements above \SI{500}{GHz}.
All experimental setups use frequency modulation with a \textit{2f}-demodulation scheme to increase the signal-to-noise ratio (SNR).
As a result, absorption features look similar to a second derivative of a Voigt profile.
The lower frequency experimental setup is described in more detail elsewhere~\cite{Zingsheim2021}.

\begin{figure}[tbh]
    \centering
    \includegraphics[width=1\linewidth]{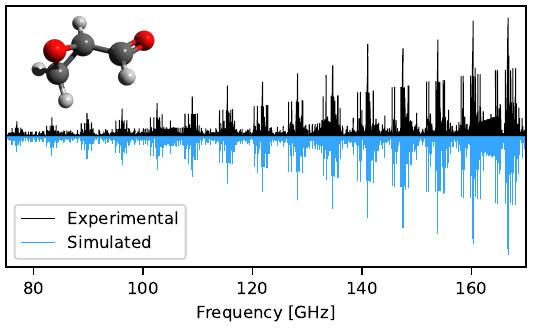}
    \caption{
    Intensity-calibrated broadband spectrum of glycidaldehyde (black) from \SI{75}{GHz} to \SI{170}{GHz} in comparison with the simulated stick spectrum (blue).
    The ground vibrational state analysis was used for the intensity calibration.
    For better visual comparability, only experimental data points with positive signals are shown here.
    The stick-and-ball representation of glycidaldehyde is shown in the top left.
    }
    \label{fig:spectrum}
\end{figure}

All measurements were performed at room temperature with filling pressures of around \SI{10}{\micro bar}.
Standing waves were removed from the measurements with a self-written Fourier filtering script\footnote{Available at \url{https://pypi.org/project/fftfilter/} or with pip via \textit{pip install fftfilter}}.
Additionally, after completing the ground vibrational state analysis, the broadband measurements were intensity calibrated using the ground vibrational state predictions (see \autoref{fig:spectrum}).
All $v=0$ transitions predicted to have an intensity $>$\SI{E-5}{nm^2 MHz} were fitted with a second derivative Voigt profile.
Blends were considered by summing the intensities of all predictions within \SI{200}{kHz} of the prediction of interest.
The ratios of the predicted intensities and fitted intensities were then used to create a calibration curve with values in between the calibration points being interpolated.
These post-processing steps greatly facilitated the visual detection of weak patterns in Loomis-Wood plots.

\subsection{Improved DM-DR Setup}
\label{sec:DMDRScheme}
An improved version of the double-modulation double-resonance (DM-DR) setup originally described in \citet{Zingsheim2021} was used for the DM-DR measurements.
The great advantage of DM-DR measurements is their ability to filter the spectrum for lines (so-called probe transitions) sharing an energy level with an already known transition, the so-called pump transition.
This greatly facilitates the assignment and analysis process as it allows filtering dense spectra for specific lines which is especially useful to unambiguously identify weak, blended and/or strongly perturbed transitions.

The DM-DR setup used here is based on the conventional \SIrange{75}{120}{GHz} absorption setup described in \autoref{sec:Experiments}.
A second, more powerful source, the pump source, is added covering the frequency range of \SIrange{70}{110}{GHz} and polarized orthogonally to the probe radiation allowing for co-spatial alignment of the probe and pump radiation via a polarizing grid.
The probe frequency is tuned to record the spectrum whereas the radiation of the pump source is fixed to the frequency of a known pump transition and not measured by the detector.
Instead, the pump radiation is used to split the energy levels of the pump transition via the Autler-Townes effect~\cite{Autler1955, Cohen-Tannoudji2008_DressedAtomApproach}.
As a result, all transitions sharing an energy level with the pump transition split symmetrically into two transitions of half intensity.
\begin{figure*}[tbh]
    \centering
    \includegraphics[width=1\linewidth]{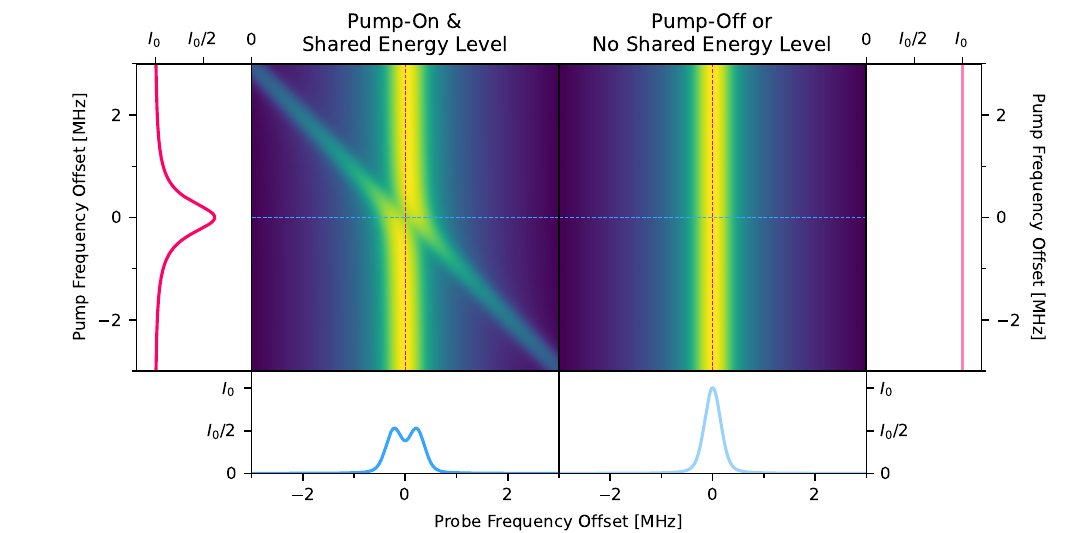}
    \caption{Simulated 2D spectra of the probe radiation absorption for the cases of a shared energy level (left half) and no pump radiation or no shared energy level between probe and pump transition (right half).
    The $x$- and $y$-axes depict the frequency offsets of the probe and pump sources, respectively.
    The dashed lines in the heatmaps indicate the slices that are shown in the plots to the sides and on the bottom.
    }
    \label{fig:pufm_setup}
\end{figure*}
This is highlighted in the simulated 2D spectra in \autoref{fig:pufm_setup}.
The left half of the figure shows the case when the probe and pump transition share an energy level and the pump radiation is on.
The heatmap shows the absorption of the probe radiation for the offsets of the probe and pump frequencies, i.e., that both sources are resonant with their respective transitions in the center of the heatmap.
The plots to the left and the bottom show the 1D spectra along the dashed lines in the same colors.
The right half shows the case when the probe and pump transitions do not share an energy level or the pump radiation is turned off.
It is apparent that in this case, the 2D spectrum does not depend on the pump frequency.

A straight-forward way to determine if the probe and pump transitions share an energy level is to subtract the spectra measured with and without pump source - basically subtracting the left and right 1D slices in blue.
The subtraction should be performed on a very short time scale to mitigate the influence of fluctuations in the experimental conditions which could otherwise lead to false positive signals (see Fig.\ 8 in \citet{Zingsheim2021}).
This was realized in the previous setup~\cite{Zingsheim2021} by amplitude modulating the pump source and a corresponding \textit{1f}-demodulation of the detector signal.
Additionally, a frequency modulation of the probe source combined with a \textit{2f}-demodulation of the detector signal was employed to increase the SNR.

The two consecutive demodulation steps are non-ideal for multiple reasons.
Firstly, using a sine signal to realize a subtraction has an efficiency of only $2/\pi \approx 64\%$.
Secondly, two lock-in amplifiers arranged in serial result in many of their parameters influencing each other (e.g.\ the dependence of the signal on both time constants).

These shortcomings can be mitigated by using a digital DM-DR setup which performs the second demodulation in the measurement software.
The intensity is measured with and without the pump source by digitally turning the pump source's radio frequency power on and off and subtracting the respective intensities on the computer. However, this procedure introduces additional overhead in the form of switching the radio frequency on and off\footnote{This overhead has greater efficiency than the $2/\pi \approx 64\%$ efficiency of the previous DM-DR setup for integration times $\geq \SI{28}{ms}$. This integration time is well surpassed in our measurements.}.

Another possibility, applied here, is to operate the probe source in continuous wave mode and apply an FM to the pump source requiring only a single lock-in amplifier for the demodulation.
The single FM of the pump source simultaneously increases the SNR and removes all transitions from the DM-DR spectrum that do not share an energy level with the pump transition.
The working principle is explained by the red 1D slices along the pump frequency axis in \autoref{fig:pufm_setup}.
If the probe and pump transition share an energy level, the signal along the pump frequency axis has a strong dip at the center position whereas if no energy level is shared the signal is constant (left-most and right-most plots of \autoref{fig:pufm_setup}, respectively).
The two cases are distinguished via the FM of the pump source and subsequent \textit{2f}-demodulation of the detector signal.
For the pump-off case, this results in a zero signal after the lock-in amplifier but a strong signal for the pump-on case.
Analogously to the conventional case, the FM also improves the SNR for the pump-on case.
However, the FM amplitude has to be adjusted as the linewidth of the conventional case (bottom right plot of \autoref{fig:pufm_setup}) depends on the different broadening effects but the linewidth along the pump frequency axis (left-most plot of \autoref{fig:pufm_setup}) also strongly depends on the magnitude of the Autler-Townes splitting.

The simplicity of the new setup is highlighted by it differing from the conventional setup only in that the lock-in amplifier is connected to the pump instead of the probe source synthesizer (and the obvious addition of a pump source).
This makes switching between the two measurement techniques as simple as changing a single cable connection.

Additionally, the current pump-modulated setup showed greater sensitivity to the detuning of the pump source in our measurements and simulations.
Therefore, false positive signals (due to transitions close to the pump transition being pumped off-resonantly) are more easily distinguished.
However, simulations also show that this behavior is very dependent on the magnitude of the Autler-Townes splitting and the chosen FM amplitudes.
In summary, for molecules with similar (transition) dipole moments as glycidaldehyde, the new setup is noticeably more sensitive to false positive signals arising from off-resonant pumping than the previous setup~\cite{Zingsheim2021}.

\section{Quantum Chemical Calculations}
\label{sec:QCC}

Complementary quantum chemical calculations of glycidaldehyde 
have been performed
at the coupled-cluster singles and doubles (CCSD) level 
augmented by a perturbative treatment of triple excitations, CCSD(T) \cite{raghavachari_chemphyslett_157_479_1989}, 
together with correlation consistent
polarized valence and polarized weighted core-valence basis
sets, as well as atomic natural orbital basis sets, specifically,
cc-pVTZ \cite{dunning_JCP_90_1007_1989}, cc-pwCVTZ \cite{peterson_JCP_117_10548_2002}, 
and ANO0 \cite{almlof_JCP_86_4070_1987}.
Equilibrium geometries have been calculated using analytic gradient techniques
\cite{watts_chemphyslett_200_1-2_1_1992}, while
harmonic frequencies have been computed using analytic second-derivative techniques
\cite{gauss_chemphyslett_276_70_1997,stanton_IntRevPhysChem_19_61_2000}.
For anharmonic computations using the cc-pVTZ and ANO0 basis sets second-order vibrational perturbation theory (VPT2) \cite{mills_alphas}   
has been employed and additional numerical differentiation of analytic second derivatives has been applied 
to obtain the third and fourth derivatives required for the application of VPT2 \cite{stanton_IntRevPhysChem_19_61_2000,stanton_JCP_108_7190_1998}.

All calculations have been carried out using 
the CFOUR program package \cite{cfour,Matthews2020}; 
for some of the calculations the parallel version of CFOUR 
\cite{harding_JChemTheoryComput_4_64_2008} 
has been used.
The resulting rotation-vibration interaction constants, harmonic and fundamental wavenumbers, fundamental intensities, energy-dependence on the aldehyde torsional angle, and optimized molecular structures are given in Sec.\ 1 of the \refsuppinfo. 

\section{Spectroscopic Fingerprint of Glycidaldehyde}
\label{sec:Glycidaldehyde}
Glycidaldehyde (\ce{(C2H3O)CHO}) is an asymmetric rotor with Ray's asymmetry parameter of $\kappa = (2B-A-C)/(A-C) =\SI{-0.98}{}$ which is very close to the prolate limit of \SI{-1}{}.
Creswell et al.\ determined its three dipole moment components to be $\mu_a = \SI{1.932(5)}{D}$, $\mu_b = \SI{1.511(17)}{D}$, and $\mu_c = \SI{0.277(156)}{D}$ resulting in strong $a$- and $b$-type spectra accompanied by a considerably weaker $c$-type spectrum~\cite{Creswell1977}.
\begin{figure}
    \centering
    \includegraphics[width=1\linewidth]{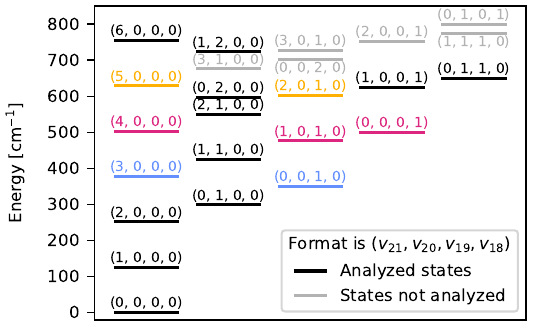}
    \caption{
    All vibrational states below \SI{800}{cm^{-1}}.
    The 18 analyzed states are shown in color or black.
    Interaction systems are indicated by mutual colors.
    }
    \label{fig:VibrationalLevels}
\end{figure}
Additionally, four fundamental vibrational modes lie below \SI{500}{cm^{-1}} resulting in a plethora of vibrationally excited states below \SI{800}{cm^{-1}} (see \autoref{fig:VibrationalLevels}).
The lowest vibrational state $\nu_{21}$, the aldehyde torsion, has an energy of about \SI{125}{cm^{-1}} only for which reason even multiply excited states up to $v_{21}=6$ could be assigned in the spectrum.
The combination of the many vibrational satellite patterns and the three non-zero dipole moments gives rise to a dense and feature-rich spectrum, making it difficult to identify weak series, especially by visual means.
For brevity, vibrational states will also be given in the format $(v_{21}, v_{20}, v_{19}, v_{18})$.

The spectroscopic assignment process relied heavily on Loomis-Wood plots (LWPs) as implemented in the LLWP software~\cite{Bonah2022} while Pickett's SPFIT and SPCAT~\cite{Pickett1991} were used for line fitting and predicting the data with an asymmetric top Hamiltonian in the 
S-reduction~\cite{Winnewisser1972} and $\text{I}^\text{r}$ representation.
Our Python library \textit{Pyckett} was used to automate repetitive tasks carried out with SPFIT and SPCAT\footnote{
Especially the interaction systems benefited from the ability to automatically test the influence of additional interaction parameters on the model via Pyckett's CLI tool \texttt{pyckett\_add}.
See \texttt{\url{https://pypi.org/project/pyckett/}} for more information or install with pip via \texttt{pip install pyckett}}.
Line uncertainties were assigned with a semi-automatic procedure described previously \cite{BonahEthylCP}.
Frequency uncertainties for the assignments by Creswell et al.\ were increased manually due to obvious systematic shifts in their data (see Sec.\ 2 of the \refsuppinfo).
Uncertainties of \SI{200}{kHz} were found appropriate were found appropriate as then the assignments could be included with negligible effects on the \textit{WRMS} value.

The quantum chemical calculations described in \autoref{sec:QCC} provided initial rotational parameters and estimates for the vibrational energies (see Sec.\ 1 of the \refsuppinfo).
The ground vibrational state and the aldehyde torsion up to $v_{21} = 6$ were straightforward to assign due to the previous microwave study~\cite{Creswell1977}.
Further vibrational satellites were identified subsequently in Loomis-Wood plots or via DM-DR measurements.
Candidates for the DM-DR measurements were found either visually in the LWPs or by an automated approach using a peakfinder.
For the latter, the identified peaks were filtered for so-far unassigned peaks which were ordered by intensity and then tested for a shared energy level via DM-DR measurements.
Furthermore, the DM-DR measurements were used to confirm and extend series showing strong deviations caused by interactions or series weak in intensity due to high vibrational energies.
Once a few series belonging to the same vibrational state were assigned, the predictions were typically sufficient to extend the assignment by LWPs into regions where no DR spectroscopy was available.

\begin{table*}[tb]

    \centering
    \caption{Approximated energies from the Boltzmann analysis $\tilde{\nu}_\text{appr}$ in \si{cm^{-1}}, rotational constant differences $\Delta A$, $\Delta B$, $\Delta C$ with respect to the ground vibrational state of the main isotopologue in \si{MHz}, the highest assigned frequency (in \si{GHz}), $J$ value, and $K_a$ value, the number of rejected lines, the number of transitions, \textit{RMS} in \si{kHz}, and \textit{WRMS} values for the analyzed states. Comparison with the calculated values (see Sec.\ 1 of the \refsuppinfo) yielded the state labels given in the first column. Interaction systems are indicated by mutual colors analog to \autoref{fig:VibrationalLevels}.}
    \label{tab:vibstates}
    \resizebox{\linewidth}{!}{
    \begin{tabular}{l S[table-format=4.3] S[table-format=-3.2] S[table-format=-3.2] S[table-format=-3.2] S[table-format=3.0] S[table-format=3.0] S[table-format=3.0] S[table-format=3.0] S[table-format=4.0] S[table-format=3.2] S[table-format=2.2]}
        \toprule
        Vib. State & \text{$\tilde{\nu}_\text{appr}$} & \text{$\Delta A$} & \text{$\Delta B$} &  \text{$\Delta C$} & \text{$\nu_\text{max}$} &  \text{$J_\text{max}$} &  \text{$K_{a,\text{max}}$} & \text{Rej.} & \text{Trans.} & \textit{RMS} & \textit{WRMS}  \\
        \midrule
$(0, 0, 0, 0)$               &   0( 0)    &     0.00 &     0.00 &     0.00 &  749 & 114 &  24 &     0       &  6921       &  28.52       & 0.86      \\
$(1, 0, 0, 0)$               & 125(27)    &  -138.84 &    12.89 &     5.16 &  749 & 115 &  24 &     0       &  5218       &  30.05       & 0.89      \\
$(2, 0, 0, 0)$               & 257(42)    &  -273.89 &    25.64 &    10.27 &  747 & 103 &  24 &     0       &  3091       &  18.84       & 0.58      \\
$(0, 1, 0, 0)$               & 312(14)    &   125.26 &    -1.82 &    -0.54 &  736 & 110 &  21 &     0       &  2356       &  17.98       & 0.59      \\
\color{blue}$(0, 0, 1, 0)$   & 361(17)    &   -17.71 &    -0.73 &    -4.72 &  737 & 100 &  20 &     1$^{a}$ &  2614$^{a}$ &  28.13$^{a}$ & 0.87$^{a}$\\
\color{blue}$(3, 0, 0, 0)$   & 390(41)    &  -406.36 &    38.33 &    15.34 &  737 & 100 &  19 &     1$^{a}$ &  2614$^{a}$ &  28.13$^{a}$ & 0.87$^{a}$\\
$(1, 1, 0, 0)$               & 437(22)    &   -23.99 &    11.48 &     4.46 &  170 &  60 &  20 &     0       &   793       &  18.14       & 0.56      \\
\color{pink}$(1, 0, 1, 0)$   & 469(20)    &  -132.47 &    10.50 &     0.32 &  170 &  52 &  21 &     0$^{b}$ &  1930$^{b}$ &  33.83$^{b}$ & 0.94$^{b}$\\
\color{pink}$(0, 0, 0, 1)$   & 516(14)    &    -3.18 &    -1.84 &    -2.01 &  168 &  51 &  19 &     0$^{b}$ &  1930$^{b}$ &  33.83$^{b}$ & 0.94$^{b}$\\
\color{pink}$(4, 0, 0, 0)$   & 538(54)    &  -533.90 &    50.87 &    20.47 &  170 &  45 &  19 &     0$^{b}$ &  1930$^{b}$ &  33.83$^{b}$ & 0.94$^{b}$\\
$(2, 1, 0, 0)$               & 584(37)    &  -166.96 &    24.45 &     9.34 &  170 &  54 &  19 &     0       &   676       &  25.11       & 0.74      \\
\color{yellow}$(2, 0, 1, 0)$ & 595(40)    &  -253.19 &    23.29 &     6.11 &  169 &  27 &  18 &     0$^{c}$ &  1016$^{c}$ &  23.74$^{c}$ & 0.66$^{c}$\\
$(0, 2, 0, 0)$               & 612(33)    &   250.42 &    -3.61 &    -1.05 &  168 &  50 &  18 &    15       &   541       &  54.72       & 1.48      \\
$(1, 0, 0, 1)$               & 637(21)    &  -148.11 &    11.91 &     3.28 &  169 &  54 &  20 &    27       &   564       &  82.60       & 2.14      \\
\color{yellow}$(5, 0, 0, 0)$ & 640(75)    &  -666.26 &    63.13 &    25.08 &  170 &  26 &  19 &     0$^{c}$ &  1016$^{c}$ &  23.74$^{c}$ & 0.66$^{c}$\\
$(0, 1, 1, 0)$               & 677(22)    &    89.94 &    -1.36 &    -4.75 &  168 &  27 &   9 &     5       &   223       &  87.91       & 2.32      \\
$(1, 2, 0, 0)$               & 754(24)    &    90.18 &    10.10 &     3.77 &  169 &  27 &  14 &     0       &   359       &  23.94       & 0.62      \\
$(6, 0, 0, 0)$               & 767(61)    &  -682.75 &    65.89 &    24.96 &  170 &  26 &   9 &   110       &   269       & 195.76       & 4.22      \\
\midrule                     
\ce{^{13}C_2} $v=0$          & 0          &  -141.82 &    -5.77 &    -1.11 &  168 &  27 &   5 &     0       &   107       &  38.70       & 0.59      \\
\ce{^{13}C_1} $v=0$          & 0          &  -333.54 &   -42.43 &   -48.41 &  168 &  27 &  12 &     0       &   354       &  30.18       & 0.59      \\
\ce{^{13}C_4} $v=0$          & 0          &   -84.48 &   -22.40 &   -18.28 &  170 &  27 &  13 &     0       &   347       &  31.07       & 0.54      \\
        \bottomrule
    \end{tabular}
    }
    \tnotes{
The rotational constant differences are defined as $\Delta A$ = $A_v$ - $A_0 \approx -\alpha^{A}_v$ (analog for $B$ and $C$). The $\alpha^{A/B/C}_v$ are the first-order rotation-vibration interaction constants.\newline
$^{a,b,c}$  Reported values are values for the respective combined fits.
}
\end{table*}

\begin{figure}[tbh]
    \centering
    \includegraphics[width=1\linewidth]{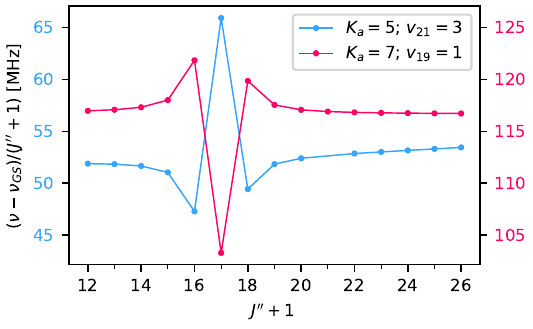}
    \caption{Resonance plot for the $3\nu_{21}$/$\nu_{19}$ dyad. A $\Delta K_a=2$ interaction for the $K_a=5$ series of $v_{21}=3$ and the $K_a=7$ series of $v_{19}=1$ is highlighted by the mirrored patterns.
    Here, the $J=K_a+K_c$ asymmetry components are shown but the figure is nearly identical for the other asymmetry component.
    The respective $y$-axes are shifted to highlight the mirror image nature of the two series which is a clear sign of an interaction between the two states centered around $J=17$.}
    \label{fig:ResonancePlot}
\end{figure}

\begin{figure*}[tb]
    \centering
    \includegraphics[width=1\linewidth]{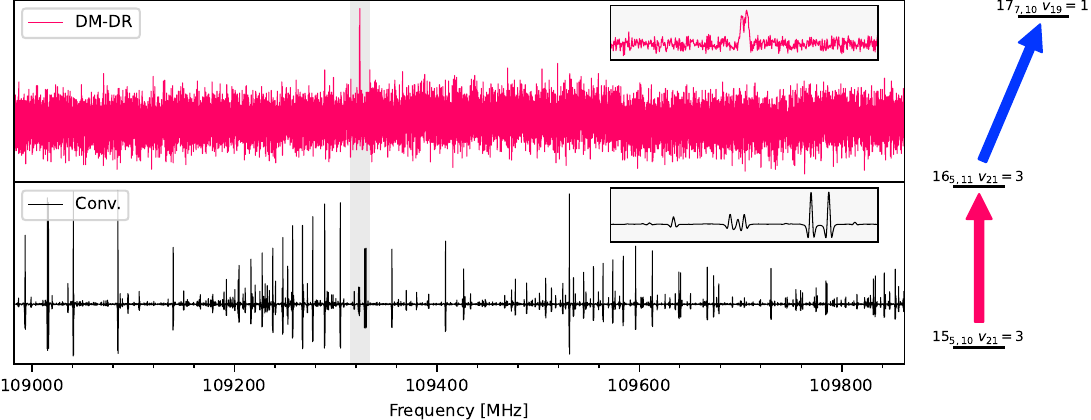}
    \caption{DM-DR (top) and conventional measurement (bottom) of the same frequency range. On the right, the transition of interest, an interstate transition, is indicated in blue and the pump transition is indicated in red. Due to the DM-DR scheme, only lines sharing an energy level with the pump transition have a signature in the top plot, unambiguously identifying the weak interstate transition. As can be seen from the zoom into the gray marked area, there are many candidate lines in the conventional spectrum but only a single line in the DM-DR spectrum.}
    \label{fig:InterstateTransition}
\end{figure*}

To assign the identified vibrationally excited states the correct labels, their rotational constant differences $\Delta A$, $\Delta B$, and $\Delta C$ (which are to first order the negative rotation-vibration interaction constants $\alpha^{A,B,C}_v$) as well as their approximated vibrational energies\footnote{The energies were approximated via a Boltzmann analysis comparing the amplitudes of the $J_{0, J} \leftarrow J-1_{0, J-1}$ transitions for $J=12$ to $J=26$ with the respective ground vibrational state transitions.
To make the results more robust, the three ratios with the highest deviation to the mean were iteratively removed.}
(see \autoref{tab:vibstates}) were compared with the values from the quantum chemical calculations.
As shown in \autoref{fig:VibrationalLevels}, the \SI{17}{} vibrationally excited states analyzed here were assigned to the \SI{15}{} energetically lowest vibrational states as well as to $(1, 2, 0, 0)$ and $(6, 0, 0, 0)$.
The experimental vibrational energies usually agree with the calculated anharmonic values within uncertainties.
The deviations are typically around \SI{10}{cm^{-1}} with a maximum deviation of \SI{56}{cm^{-1} (\SI{1.04}{\sigma})} for $(4, 0, 0, 0)$.
The rejected lines and \textit{RMS} values in \autoref{tab:vibstates} show that the vibrational states $(0, 2, 0, 0)$, $(1, 0, 0, 1)$, $(0, 1, 1, 0 )$, and $(6, 0, 0 ,0)$\footnote{The vibrational state $(6, 0 , 0 , 0)$ is most likely interacting with the not yet found $(3, 0, 1, 0)$ state, as $(3+n, 0, 0, 0)$ and $(n, 0, 1, 0)$ are interaction partners for $n=0,1,2$. Similarly, $(1, 0, 0, 1)$ could be interacting with $(5, 0, 0, 0)$ and $(2, 0, 1, 0)$. However, no conclusive evidence was found in the present analysis.}
could not be fit to experimental accuracy, hinting toward so far unaccounted interactions between these states and/or states not found yet.
Therefore, these vibrational states were only fit to (incomplete) sets of quartic parameters as higher-order parameters had physically unreasonable values (see Sec.\ 3 of the \refsuppinfo).
The obtained quartic distortion constants are probably highly effective and therefore should be viewed with caution.
Lastly, we searched for the gauche rotamer which is calculated to lie about \SI{2.4}{kcal/mol} above the trans rotamer which is studied here (see Sec.\ 1.4 of the \refsuppinfo).
However, it could not be identified in the spectrum due to its low relative intensity of about \SI{1}{\percent}.

\begin{table*}[tbp]
\caption{Rotational parameters for the vibrationally excited states showing no signs of interactions.}
\label{tab:RotParameters}\resizebox{\linewidth}{!}{
\begin{tabular}{l l *{ 7 }{S[table-format=-5.7]}}
\toprule
\multicolumn{2}{l}{Parameter} & \text{$(0, 0, 0, 0)$} & \text{$(1, 0, 0, 0)$} & \text{$(2, 0, 0, 0)$} & \text{$(0, 1, 0, 0)$} & \text{$(1, 1, 0, 0)$} & \text{$(2, 1, 0, 0)$} & \text{$(1, 2, 0, 0)$} \\
\midrule
$ A          $&/$ \si{\mega\hertz}     $& 18241.10399(19)& 18102.25986(19)& 17967.21715(20)& 18366.36399(22)& 18217.11177(70)&  18074.1420(16)&  18331.2800(78)\\ 
$ B          $&/$ \si{\mega\hertz}     $& 3272.930493(18)& 3285.825194(21)& 3298.572744(22)& 3271.112666(24)& 3284.406457(47)&  3297.37747(11)&  3283.02558(25)\\ 
$ C          $&/$ \si{\mega\hertz}     $& 3137.715975(18)& 3142.876045(20)& 3147.986349(21)& 3137.177292(23)& 3142.173621(47)&  3147.05316(10)&  3141.48199(23)\\ 
$ -D_{J}     $&/$ \si{\hertz}          $&   -562.8140(33)&   -584.9656(39)&   -609.5946(73)&   -558.7139(73)&    -579.461(37)&    -603.725(99)&     -574.30(13)\\ 
$ -D_{JK}    $&/$ \si{\kilo\hertz}     $&   -12.38466(13)&  -11.719947(89)&   -11.02724(14)&   -12.72937(18)&   -12.04424(30)&    -11.3339(11)&    -12.3501(24)\\ 
$ -D_{K}     $&/$ \si{\kilo\hertz}     $&   -17.02832(89)&   -13.88504(70)&   -11.65620(98)&    -21.0938(12)&     -16.643(31)&     -14.020(72)&        \text{a}\\ 
$ d_{1}      $&/$ \si{\hertz}          $&   -23.78544(78)&    -28.8834(11)&    -35.1889(15)&    -21.8177(19)&    -25.6245(23)&    -31.1607(65)&      -22.27(13)\\ 
$ d_{2}      $&/$ \si{\hertz}          $&    -3.80778(59)&    -4.65205(34)&    -5.65533(47)&    -4.10739(18)&    -4.84136(64)&     -5.8702(38)&      -4.952(88)\\ 
$ H_{J}      $&/$ \si{\micro\hertz}    $&      135.01(19)&      180.34(23)&      225.13(70)&      154.89(59)&        \text{a}&        \text{a}&        \text{a}\\ 
$ H_{JK}     $&/$ \si{\milli\hertz}    $&      74.081(31)&      70.101(11)&      65.710(21)&      76.983(33)&        \text{a}&       60.39(47)&        \text{a}\\ 
$ H_{KJ}     $&/$ \si{\hertz}          $&    -1.11992(26)&    -1.02802(29)&    -0.95695(62)&    -1.17080(94)&     -1.0776(11)&     -1.1700(26)&      -1.440(14)\\ 
$ H_{K}      $&/$ \si{\hertz}          $&      0.9334(23)&     0.63058(86)&      0.4280(18)&      1.0528(23)&        \text{a}&        \text{a}&        \text{a}\\ 
$ h_{1}      $&/$ \si{\micro\hertz}    $&      66.762(59)&      86.997(91)&      103.90(21)&       80.33(24)&        \text{a}&        \text{a}&        \text{a}\\ 
$ h_{2}      $&/$ \si{\micro\hertz}    $&       65.81(19)&      74.300(59)&       80.97(12)&        \text{a}&        \text{a}&        \text{a}&        \text{a}\\ 
$ h_{3}      $&/$ \si{\micro\hertz}    $&       3.146(17)&       4.356(23)&       5.788(29)&       4.285(26)&        \text{a}&        \text{a}&        \text{a}\\ 
$ L_{JJK}    $&/$ \si{\nano\hertz}     $&     -431.2(2.1)&        \text{a}&        \text{a}&        \text{a}&        \text{a}&        \text{a}&        \text{a}\\ 
$ L_{JK}     $&/$ \si{\micro\hertz}    $&      15.372(66)&        \text{a}&        \text{a}&        \text{a}&        \text{a}&        \text{a}&        \text{a}\\ 
$ L_{KKJ}    $&/$ \si{\micro\hertz}    $&     -148.35(20)&     -138.79(38)&     -119.3(1.4)&     -195.9(2.4)&        \text{a}&        \text{a}&        \text{a}\\ 
$ L_{K}      $&/$ \si{\micro\hertz}    $&       91.6(2.1)&        \text{a}&        \text{a}&        \text{a}&        \text{a}&        \text{a}&        \text{a}\\ 
$ l_{2}      $&/$ \si{\pico\hertz}     $&        -456(12)&        \text{a}&        \text{a}&        \text{a}&        \text{a}&        \text{a}&        \text{a}\\ 
$ P_{JK}     $&/$ \si{\pico\hertz}     $&     -182.4(5.1)&        \text{a}&        \text{a}&        \text{a}&        \text{a}&        \text{a}&        \text{a} \\
\bottomrule
\end{tabular}
}
\tnotes{
Fits performed with SPFIT in the S-reduction and $\text{I}^\text{r}$ representation.
Standard errors are given in parentheses.
$^a$ Parameter was fixed to the ground vibrational state value.
}
\end{table*}

\begin{table*}[tbp]
\caption{Rotational parameters for perturbed vibrationally excited states. Interaction systems are indicated by mutual colors analog to \autoref{fig:VibrationalLevels}.}
\label{tab:RotParametersInteractionSystems}
\resizebox{\linewidth}{!}{
\begin{tabular}{l l *{ 7 }{S[table-format=-5.7]}}
\toprule
\multicolumn{2}{l}{Parameter} & \text{\color{blue}$(3, 0, 0, 0)$} & \text{\color{blue}$(0, 0, 1, 0)$} & \text{\color{pink}$(4, 0, 0, 0)$} & \text{\color{pink}$(0, 0, 0, 1)$} & \text{\color{pink}$(1, 0, 1, 0)$} & \text{\color{yellow}$(5, 0, 0, 0)$} & \text{\color{yellow}$(2, 0, 1, 0)$} \\
\midrule
$ A          $&/$ \si{\mega\hertz}     $& 17834.74075(93)& 18223.39307(75)&   17707.207(32)&  18237.9197(18)&   18108.639(31)&   17574.848(64)&   17987.917(50)\\ 
$ B          $&/$ \si{\mega\hertz}     $&  3311.25931(48)&  3272.20455(49)&   3323.7963(60)&  3271.08786(14)&   3283.4264(60)&   3336.0643(43)&   3296.2156(44)\\ 
$ C          $&/$ \si{\mega\hertz}     $&  3153.05977(45)&  3132.99173(47)&   3158.1879(37)&  3135.70694(14)&   3138.0404(41)&   3162.8007(46)&   3143.8252(44)\\ 
$ -D_{J}     $&/$ \si{\hertz}          $&    -635.605(28)&    -567.669(36)&     -683.43(49)&     -551.24(13)&     -582.05(47)&     -687.06(40)&     -632.22(34)\\ 
$ -D_{JK}    $&/$ \si{\kilo\hertz}     $&   -10.33109(79)&   -12.19693(61)&     -9.2939(43)&   -12.87107(74)&    -11.6037(31)&     -8.8172(96)&     -10.777(10)\\ 
$ -D_{K}     $&/$ \si{\kilo\hertz}     $&     -9.9572(27)&    -18.3391(19)&       -9.44(11)&     -11.235(85)&     -19.105(83)&       -5.3(2.2)&       -3.6(1.8)\\ 
$ d_{1}      $&/$ \si{\hertz}          $&     -42.790(10)&    -29.1396(97)&      -54.78(12)&    -14.4779(93)&      -39.04(11)&      -56.96(24)&      -51.52(21)\\ 
$ d_{2}      $&/$ \si{\hertz}          $&     -6.7420(23)&     -4.2186(20)&      -9.753(34)&      -1.847(14)&      -5.786(29)&       -9.18(12)&       -7.25(10)\\ 
$ H_{J}      $&/$ \si{\micro\hertz}    $&        \text{a}&      198.2(1.8)&        \text{a}&        \text{a}&        \text{a}&        \text{a}&        \text{a}\\ 
$ H_{JK}     $&/$ \si{\milli\hertz}    $&       62.31(22)&        \text{a}&        \text{a}&        \text{a}&        \text{a}&        \text{a}&       35.7(1.8)\\ 
$ H_{KJ}     $&/$ \si{\hertz}          $&     -0.8813(11)&        \text{a}&        \text{a}&     -0.8042(31)&     -0.9352(48)&     -0.6181(59)&        \text{a}\\ 
$ H_{K}      $&/$ \si{\milli\hertz}    $&      330.0(5.3)&        \text{a}&        \text{a}&        \text{a}&        \text{a}&        \text{a}&        \text{a}\\ 
$ h_{2}      $&/$ \si{\micro\hertz}    $&      100.30(93)&        \text{a}&        \text{a}&        \text{a}&        \text{a}&        \text{a}&        \text{a}\\ 
\bottomrule
\end{tabular}
}
\tnotes{
Fits performed with SPFIT in the S-reduction and $\text{I}^\text{r}$ representation.
Standard errors are given in parentheses.
Parameters not shown here were fixed to the ground vibrational state values (see \autoref{tab:RotParameters}).
$^a$ Parameter was fixed to the ground vibrational state value.
}
\end{table*}

\begin{table}[tb]
\caption{Energy differences and interaction parameters for the three interacting systems. Interaction systems are ordered by increasing energy and are indicated by mutual colors analog to \autoref{fig:VibrationalLevels}.}
\label{tab:InteractionParameters}
\resizebox{\linewidth}{!}{
\begin{tabular}{l l r l l S[table-format=4.9]}
\toprule
\multicolumn{1}{c}{$v_1$} & \multicolumn{1}{c}{$v_2$} & \multicolumn{1}{c}{ID$^a$} & \multicolumn{2}{c}{\text{Parameter}} & \text{Value} \\
\midrule
\multicolumn{4}{r}{{\color{blue}$\tilde{\nu}_{(3, 0, 0, 0)} - \tilde{\nu}_{(0, 0, 1, 0)}$}} & /\si{cm^{-1}}  & 11.244843(37) \\
$(3, 0, 0, 0)$  & $(0, 0, 1, 0)$  &      0$v_1v_2$ & $W$             & /\si{\giga\hertz}     &         -65.6207(14) \\
$(3, 0, 0, 0)$  & $(0, 0, 1, 0)$  &      1$v_1v_2$ & $W_{J}$         & /\si{\kilo\hertz}     &           295.1(1.2) \\
$(3, 0, 0, 0)$  & $(0, 0, 1, 0)$  &    400$v_1v_2$ & $W_{2}$         & /\si{\kilo\hertz}     &          -28.477(63) \\
$(3, 0, 0, 0)$  & $(0, 0, 1, 0)$  &    401$v_1v_2$ & $W_{2,J}$       & /\si{\hertz}          &           -1.848(22) \\
$(3, 0, 0, 0)$  & $(0, 0, 1, 0)$  &   4000$v_1v_2$ & $G_{b}$         & /\si{\mega\hertz}     &          -164.86(29) \\
$(3, 0, 0, 0)$  & $(0, 0, 1, 0)$  &   6000$v_1v_2$ & $G_{c}$         & /\si{\mega\hertz}     &           177.28(27) \\
$(3, 0, 0, 0)$  & $(0, 0, 1, 0)$  &   6001$v_1v_2$ & $G_{c,J}$       & /\si{\kilo\hertz}     &           -1.684(18) \\
$(3, 0, 0, 0)$  & $(0, 0, 1, 0)$  &   6100$v_1v_2$ & $F_{ab}$        & /\si{\mega\hertz}     &           -2.585(15) \\
\midrule
\multicolumn{4}{r}{{\color{pink}$\tilde{\nu}_{(0, 0, 0, 1)} - \tilde{\nu}_{(4, 0, 0, 0)}$}} & /\si{cm^{-1}}  &  25.8639(64) \\
$(4, 0, 0, 0)$  & $(0, 0, 0, 1)$  &   2100$v_1v_2$ & $F_{bc}$        & /\si{\mega\hertz}     &            1.958(11) \\
$(4, 0, 0, 0)$  & $(0, 0, 0, 1)$  &   2101$v_1v_2$ & $F_{bc,J}$      & /\si{\hertz}          &          -128.2(7.5) \\
\multicolumn{4}{r}{{\color{pink}$\tilde{\nu}_{(4, 0, 0, 0)} - \tilde{\nu}_{(1, 0, 1, 0)}$}} & /\si{cm^{-1}}  &  12.6795(50) \\
$(4, 0, 0, 0)$  & $(1, 0, 1, 0)$  &      0$v_1v_2$ & $W$             & /\si{\giga\hertz}     &           134.42(10) \\
$(4, 0, 0, 0)$  & $(1, 0, 1, 0)$  &     10$v_1v_2$ & $W_{K}$         & /\si{\mega\hertz}     &            -9.71(14) \\
$(4, 0, 0, 0)$  & $(1, 0, 1, 0)$  &   4000$v_1v_2$ & $G_{b}$         & /\si{\mega\hertz}     &           422.7(1.1) \\
$(4, 0, 0, 0)$  & $(1, 0, 1, 0)$  &   4010$v_1v_2$ & $G_{b,K}$       & /\si{\mega\hertz}     &           -1.250(17) \\
$(4, 0, 0, 0)$  & $(1, 0, 1, 0)$  &   4200$v_1v_2$ & $G_{2b}$        & /\si{\kilo\hertz}     &           -4.421(77) \\
$(4, 0, 0, 0)$  & $(1, 0, 1, 0)$  &   4210$v_1v_2$ & $G_{2b,K}$      & /\si{\hertz}          &            53.7(1.7) \\
$(4, 0, 0, 0)$  & $(1, 0, 1, 0)$  &   4100$v_1v_2$ & $F_{ac}$        & /\si{\mega\hertz}     &            -6.91(12) \\
\midrule
\multicolumn{4}{r}{{\color{yellow}$\tilde{\nu}_{(5, 0, 0, 0)} - \tilde{\nu}_{(2, 0, 1, 0)}$}} & /\si{cm^{-1}}  &  12.9327(37) \\
$(5, 0, 0, 0)$  & $(2, 0, 1, 0)$  &      0$v_1v_2$ & $W$             & /\si{\giga\hertz}     &         -198.920(58) \\
$(5, 0, 0, 0)$  & $(2, 0, 1, 0)$  &    400$v_1v_2$ & $W_{2}$         & /\si{\kilo\hertz}     &            26.38(95) \\
$(5, 0, 0, 0)$  & $(2, 0, 1, 0)$  &   4000$v_1v_2$ & $G_{b}$         & /\si{\mega\hertz}     &          432.455(39) \\
$(5, 0, 0, 0)$  & $(2, 0, 1, 0)$  &   4101$v_1v_2$ & $F_{ac,J}$      & /\si{\hertz}          &           313.9(2.2) \\

\bottomrule
\end{tabular}
}
\tnotes{$^a$ The specified IDs are the respective parameter IDs used in the \textit{*.par} and \textit{*.var} files of SPFIT and SPCAT~\cite{Pickett1991, Drouin2017}.}
\end{table}

\begin{table}[tb]
    \caption{Rotational parameters for the ground vibrational states of the three singly substituted \ce{^{13}C} isotopologues of glycidaldehyde.}
    \label{tab:13C_Parameters}
    \resizebox{\linewidth}{!}{
    \begin{tabular}{l l *{ 3 }{S[table-format=5.9]}}
    \toprule
\multicolumn{2}{l}{Parameter} & \text{\ce{^{13}C_2} } & \text{\ce{^{13}C_1} } & \text{\ce{^{13}C_4} } \\
\midrule
$ A          $&/$ \si{\mega\hertz}     $&   18099.286(30)&   17907.566(26)&   18156.623(12)\\ 
$ B          $&/$ \si{\mega\hertz}     $&  3267.16409(56)&  3230.49876(47)&  3250.52951(16)\\ 
$ C          $&/$ \si{\mega\hertz}     $&  3136.60101(42)&  3089.30908(45)&  3119.43923(15)\\ 
$ -D_{J}     $&/$ \si{\hertz}          $&     -559.12(21)&     -554.49(12)&     -561.65(11)\\ 
$ -D_{JK}    $&/$ \si{\kilo\hertz}     $&    -12.1451(91)&   -11.91466(91)&   -11.90003(72)\\ 
$ -D_{K}     $&/$ \si{\kilo\hertz}     $&        \text{a}&        \text{a}&        \text{a}\\ 
$ d_{1}      $&/$ \si{\hertz}          $&      -22.16(24)&      -26.93(21)&        \text{a}\\ 
$ d_{2}      $&/$ \si{\hertz}          $&        \text{a}&        \text{a}&        \text{a}\\ 
\bottomrule
\end{tabular}
}
\tnotes{
Fits performed with SPFIT in the S-reduction and $\text{I}^\text{r}$ representation.
Standard errors are given in parentheses.
Parameters not shown here were fixed to the main isotopologue values (see \autoref{tab:RotParameters}).
$^a$ Parameter was fixed to the main isotopologue value.
}
\end{table}

\subsection{Interactions}
Three interacting systems, two dyads and one triad (see \autoref{fig:VibrationalLevels}), were identified via mirror images in their resonance plots (see \autoref{fig:ResonancePlot} for an example case).
For the $(4, 0, 0, 0)$, $(0, 0, 0, 1)$, and $(1, 0, 1, 0)$ triad as well as the $(5, 0, 0, 0)$, and $(2, 0, 1, 0)$ dyad, the initial vibrational energy separations were approximated from the rotational energies of the strongest mirror images in the resonance plots\footnote{
To approximate the initial vibrational energy separations, first, the energy levels at the center of the interaction are identified in resonance plots (e.g.\ $17_{7, 10}$ of $v_{19}=1$ and $17_{5, 12}$ of $v_{21}=3$ in \autoref{fig:ResonancePlot}).
Their rovibrational energies, consisting of a calculated vibrational energy and the rotational energy from the best-fit rotational model, are retrieved from the \textit{*.egy} file produced by SPCAT~\cite{Pickett1991}.
Then, the rovibrational energies of the two states are equalized by adding an energy offset to one of the two vibrational states (in the example of $v_{21}=3$ and $v_{19}=1$ the calculated vibrational energies $E^\text{calc}_{v_{19}=1}$ and $E^\text{calc}_{v_{21}=3}$ are used together with the offset energy $\Delta E_{v_{21}=3}$).
Once first interaction parameters are included in the combined fit, the offset energy can be floated in the fit and will result in a highly accurate value for the vibrational energy separation.
}.
The analysis of the $(3, 0, 0, 0)$ and $(0, 0, 1, 0)$ dyad was greatly simplified by finding interstate transitions between vibrationally excited states using the DM-DR method.
This yielded their vibrational energy separation with rotational precision (see \autoref{fig:InterstateTransition} for an example).
Transitions between vibrationally excited states become allowed due to the interactions which result in the rovibrational levels being linear combinations of the rovibrational levels of the interacting vibrational states.
This means the rovibrational levels shown on the right-hand side of \autoref{fig:InterstateTransition} are linear combinations with the labels corresponding to the largest contribution.
For example, the rovibrational level labeled $17_{7, 10} \ v_{19}=1$ is a linear combination of $17_{7, 10} \ v_{19}=1$ and $17_{5, 12} \ v_{21}=3$ (similarly $16_{5, 11} \ v_{21}=3$ will also contain parts of $16_{7, 9} \ v_{19}=1$).
As a result, the transition between the states labeled $16_{5, 11} \ v_{21}=3$ and $17_{7, 10} \ v_{19}=1$ becomes allowed. 

As glycidaldehyde has $C_1$ symmetry, $a$-, $b$-, and $c$-type Coriolis interactions as well as Fermi resonances are allowed by symmetry considerations between any combination of vibrational modes.
For each interaction system, the leading parameters for the different interaction types were tested with \textit{Pyckett} and the influence on the \textit{RMS}, \textit{WRMS}, and rejected lines were monitored.
When lower-order parameters were added, subsequently higher-order parameters of the same interaction type were tested.
The final interaction parameter set includes Fermi, $a$-, $b$-, and $c$-type Coriolis terms.
The absolute signs of the interaction parameters could not be determined from the fit, only their relative signs~\cite{Bonah2024,Islami1996}.

To find the regions where the interactions have the strongest influence, the shifts in transition frequency between predictions with and without interactions were calculated (see Sec.\ 4 of the \refsuppinfo).
When comparing these plots for coupled vibrational states, mirror images highlight the rotational states that are interaction partners.
For all three interacting systems the trends were very similar as $\Delta K_a$ decreases with increasing $K_a$ value.
Exemplary for the energetically lowest dyad, there are clear mirror image patterns for $K_a=2$ of $(3, 0, 0, 0)$ and $K_a=6$ of $(0, 0, 1, 0)$ resulting in $\Delta K_a=4$ for these lower $K_a$ values.
For higher $K_a$ values mirror image patterns can be seen e.g. for $K_a=11$ and $K_a=12$, respectively, resulting in $\Delta K_a=1$ for these higher $K_a$ values.

For low $K_a$ values the interactions between $(3, 0, 0, 0)$ and $(0, 0, 1, 0)$ are most prominent around $J=16$ for $K_a=5$ and $K_a=7$, respectively (see \autoref{fig:ResonancePlot}). 
Further strong interactions are found around $J=32$ between $K_a=11$ and $K_a=12$ in addition to $\Delta K_a=0$ interactions for the highest measured values of $K_a$, being 15 to 20.

For the $(4, 0, 0, 0)$, $(0, 0, 0, 1)$, and $(1, 0, 1, 0)$ triad, the dominant shifts are between $(4, 0, 0, 0)$ and $(1, 0, 1, 0)$. At low $K_a$ values, the prominent interactions are around $J=32$ between $K_a=7$ and $K_a=9$. For higher $K_a$ values $\Delta K_a = 1$ interactions dominate with the biggest shifts around $J=37$ between $K_a=14$ and $K_a=15$, respectively.

For the $(5, 0, 0, 0)$, and $(2, 0, 1, 0)$ dyad the strongest interactions are around $J\geq 50$ between $K_a=17$ and $K_a=18$, respectively.

Due to the limitations in the frequency coverage, $a$-type transitions of the interacting systems are typically assigned up to $J = 26$.
As a consequence, the centers of some perturbations lie outside our quantum number range and are determined via the predictions with and without interaction parameters.

The resulting rotational parameters (see \autoref{tab:RotParametersInteractionSystems}) are deemed to be effective, particularly for the $(4, 0, 0, 0)$, $(0, 0, 0, 1)$, and $(1, 0, 1, 0)$ triad, as can be seen in the parameter progression for the quartic parameters (see Sec.\ 5 of the \refsuppinfo).
This could result from so far unaccounted interactions with other states where the effects are too weak to be identified in the residuals or the corresponding quantum number ranges have not been observed yet.

Lastly, both $(2, 1, 0, 0)$, and $(1, 2, 0, 0)$ show hints for perturbations at higher $K_a$ values but no interaction partners were identified as of yet.

\subsection{Fit Results}
Except for the aforementioned vibrationally excited states,
$(0, 2, 0, 0)$, $(1, 0, 0, 1)$, $(0, 1, 1, 0 )$, and $(6, 0, 0 ,0)$, all analyses reproduce the spectrum with about experimental uncertainty and only a single line was rejected from the fit due to $|\nu_\text{obs} - \nu_\text{calc}| / \Delta \nu > 10$ (see \autoref{tab:vibstates}).
The resulting rotational parameters are given in \autoref{tab:RotParameters} for vibrational states showing no signs of interactions, in \autoref{tab:RotParametersInteractionSystems} for the interacting vibrational states, and the interaction parameters in \autoref{tab:InteractionParameters}.

In their previous study, Creswell et al.\ presented fits for $v_{21}=0, 1, 2$ with full sets of quartic rotational constants.
The relative deviations for the rotational constants are all below $10^{-5}$ even though not within their uncertainties.
The quartic parameters differ by a few percent except $d_{2}$ which differs by up to 45\%.
For $v_{21}=3,4,5,6$ Creswell et al.\ fitted only rigid rotor Hamiltonians without accounting for any interactions.
Thus, even the rotational constants differ greatly (up to $150 \sigma$\footnote{The values for $v_{21}=6$ agree the best as in this work the interaction partner of $v_{21}=6$ was not found meaning two highly effective fits were compared.}).
However, their study was much more limited in frequency coverage, only \SIrange{8}{40}{GHz}, and quantum number coverage, with only 31 transitions with $J_\text{max} = 31$ for $v_{21}=0, 1, 2$ and 10 transitions with $J_\text{max} = 6$ for $v_{21}=3, 4, 5, 6$.
Furthermore, no interactions were accounted for at all by \citet{Creswell1977}.

The values for the ground vibrational state also show good agreement with values obtained from quantum chemical calculations (see Tab.\ S2 of the \refsuppinfo).
The three parameters $H_J$, $h_1$, and $d_2$ show the highest relative deviations at \SI{27}{\percent}, \SI{26}{\percent}, and \SI{21}{\percent}, respectively.

\subsection{Singly substituted \ce{^13C} Species}
The singly substituted \ce{^13C} species were more difficult to analyze because of their low natural abundance resulting in a smaller number of assignable lines.
This is especially apparent for the \ce{^{13}C2} species, for which only about 100 lines could be assigned due to its $A$ and $B$ values being close to the main isotopologue values (see \autoref{tab:vibstates}). 
As a result, many of its $a$-type transitions are close to or blended with much stronger transitions.
The resulting parameters are shown in \autoref{tab:13C_Parameters} with the notation from Creswell et al.~\cite{Creswell1977} being used to label the carbon atoms.

The rotational constants agree nicely with the values obtained by \citet{Creswell1977} as the relative deviations are below $10^{-5}$ and all deviations are within $1.5 \sigma$.
Higher order parameters were included by \citet{Creswell1977} only as fixed values of the main isotopologue's ground vibrational state.

\subsection{Estimation of the $c$-type Dipole Moment}
Creswell et al.\ determined the $c$-type dipole moment to be \SI{0.277(156)}{D}.
To derive a more accurate value, in the present study, the $c$-type dipole moment was determined via comparison of experimental line intensities.
A selection of \SI{28}{} $c$-type transitions was compared with $b$-type transitions close in frequency by fitting a second derivative Voigt profile to the experimental lineshapes.
The resulting $c$-type dipole moment, calculated as the mean value and the standard deviation of the \SI{28}{} transition pairs, is \SI{0.334(39)}{D}.
Additionally, the calculated equilibrium $c$-type dipole moment (ae-CCSD(T)/cc-pwCVTZ level, \SI{0.276}{D}) was corrected for effects of zero-point vibrations (fc-CCSD(T)/cc-pVTZ, \SI{0.023}{D}) resulting in a vibrationally averaged dipole moment of $\mu_c = \SI{0.253}{D}$, which is about $2\sigma$ smaller than the experimental value.

The new experimental value agrees with the previously determined value within their uncertainties but has a significantly smaller uncertainty.
Systematic errors occur as (among other things) the power of the probe source is frequency dependent, and the pressure, which influences the lineshape and thereby center intensities, rises over time.
These effects were not taken into account but their impact was minimized by comparing transitions nearby in frequency (and thereby also in time).
Additionally, lines might be blended with other lines so-far not assigned and therefore their amplitudes might be influenced.
The systematic uncertainty might thus be slightly higher than the reported statistical uncertainty.

\section{Search for Glycidaldehyde toward Sgr~B2(N)}
\label{s:sgrb2}

\subsection{Observations}
\label{ss:obs_remoca}

We used the imaging spectral line survey Reexploring Molecular Complexity
with ALMA (ReMoCA) that targeted the high-mass star-forming protocluster 
Sgr~B2(N) with the Atacama Large Millimeter/submillimeter Array (ALMA) to search for glycidaldehyde in the interstellar medium. 
Details about the data reduction and the method of analysis of this survey can
be found elsewhere \cite{Belloche19,Belloche22}. The main features of 
the survey are the following. It covers the frequency range from 84.1~GHz to 
114.4~GHz at a spectral resolution of 488~kHz (1.7 to 1.3~km~s$^{-1}$). This 
frequency coverage was obtained with five different tunings of the receivers.
The phase center was located halfway between the two hot molecular cores 
Sgr~B2(N1) and Sgr~B2(N2), at the equatorial position 
($\alpha, \delta$)$_{\rm J2000}$= 
($17^{\rm h}47^{\rm m}19{\fs}87, -28^\circ22'16 {\farcs} 0$).
The observations achieved a sensitivity per spectral channel ranging between 
0.35~mJy~beam$^{-1}$ and 1.1~mJy~beam$^{-1}$ (rms) depending on the tuning, with 
a median value of 0.8~mJy~beam$^{-1}$. The angular resolution (HPBW) varies 
between $\sim$\SI{0.3}{\arcsecond} and $\sim$\SI{0.8}{\arcsecond} with a median value of 
\SI{0.6}{\arcsecond} that corresponds to $\sim$4900~au at the distance of Sgr~B2 (8.2~kpc)\cite{Reid19}. 

For this work we analyzed the spectra toward the position called Sgr~B2(N2b) 
by Belloche et al.~\cite{Belloche22}. It is located in the secondary hot core Sgr~B2(N2) at 
($\alpha, \delta$)$_{\rm J2000}$= 
($17^{\rm h}47^{\rm m}19{\fs}83, -28^\circ22'13{\farcs}6$). This position was 
chosen as a compromise between getting narrow line widths to reduce the level
of spectral confusion and keeping a high enough H$_2$ column density to detect 
less abundant molecules.

Like in our previous ReMoCA studies~\cite{Belloche19,Belloche22}, we 
compared the observed spectra to synthetic spectra computed under the 
assumption of local thermodynamic equilibrium (LTE) with the astronomical 
software Weeds \cite{Maret11}. This assumption is justified by the high 
densities of the regions where hot-core emission is detected in Sgr~B2(N) $>1 \times 10^{7}$~cm$^{-3}$, see
\citet{Bonfand19}. The calculations take
into account the finite angular resolution of the observations and the optical 
depth of the rotational transitions. We derived by eye a 
best-fit synthetic spectrum for each molecule separately, and then added 
together the contributions of all identified molecules. Each species was 
modeled with a set of five parameters: size of the emitting region 
($\theta_{\rm s}$), column density ($N$), temperature ($T_{\rm rot}$), linewidth 
($\Delta V$), and velocity offset ($V_{\rm off}$) with respect to the assumed 
systemic velocity of the source, $V_{\rm sys}=74.2$~km~s$^{-1}$. The linewidth 
and velocity offset were obtained directly from the well-detected and not 
contaminated lines. The size of the emission of a given molecule was estimated 
from integrated  intensity maps of transitions of this molecule that were 
found to be relatively free of contamination from other species.

\subsection{Search for glycidaldehyde toward Sgr~B2(N2b)}
\label{ss:result_remoca}

\begin{figure}[!h]
\centerline{\resizebox{1.0\hsize}{!}{\includegraphics[angle=0]{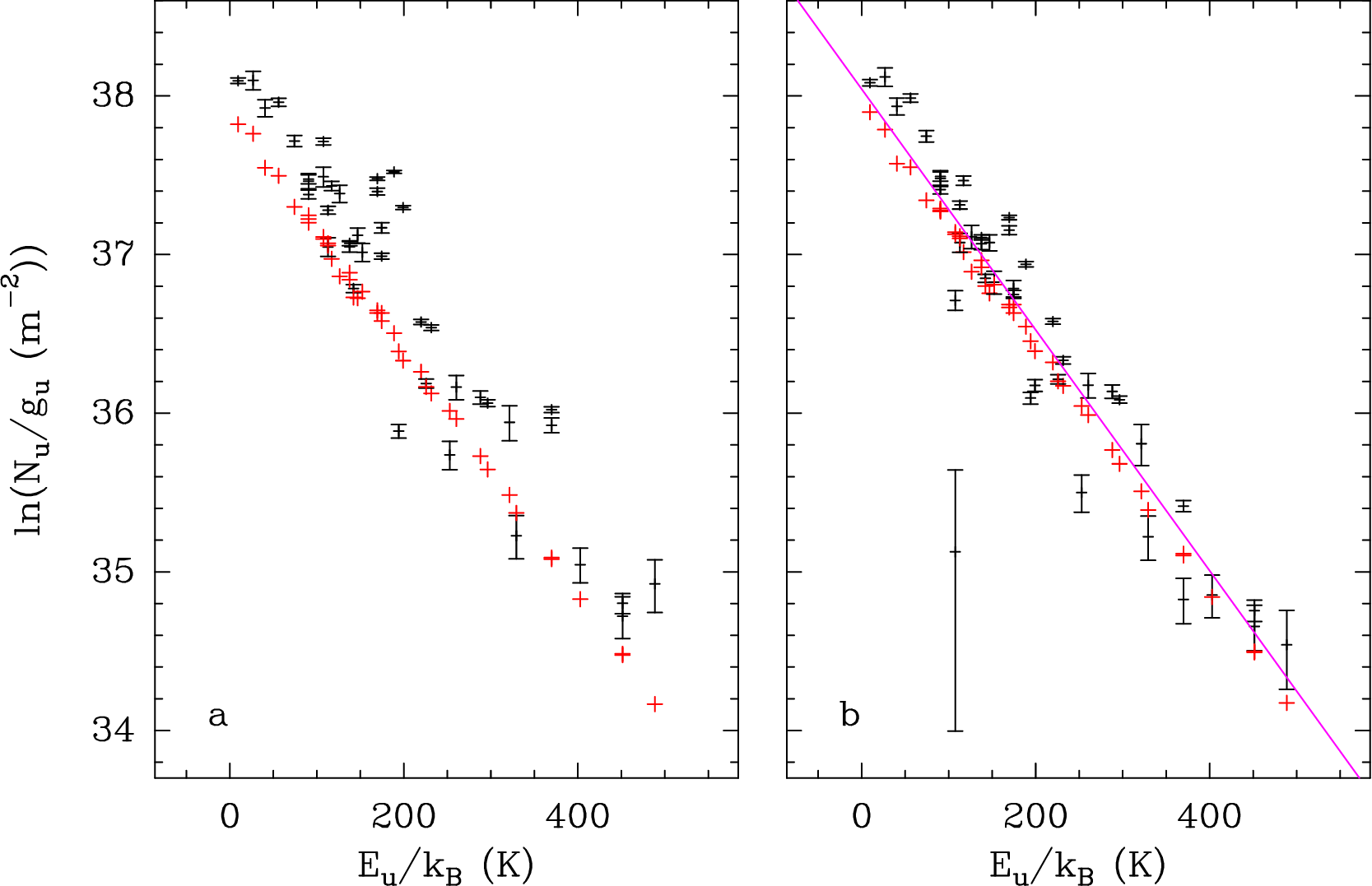}}}
\caption{Population diagram of \ce{\textit{c}-C2H4O} toward Sgr~B2(N2b). The
observed data points are shown in black while the synthetic
populations are shown in red. No correction is applied in panel {\bf a}.
In panel {\bf b}, the optical depth correction has been applied to both the
observed and synthetic populations and the contamination by all other
species included in the full model has been subtracted from the observed
data points. The purple line is a linear fit to the observed populations (in
linear-logarithmic space). The fit yields a temperature of 132 $\pm$ 10~K.
}
\label{f:popdiag_c2h4o-c_n2b}
\end{figure}

Before searching for glycidaldehyde toward Sgr~B2(N2b), we modeled the 
rotational emission of the related cyclic molecule oxirane, \ce{\textit{c}-C2H4O}. We 
used the spectroscopic predictions available in the Cologne database for 
molecule spectroscopy (CDMS)~\cite{Mueller05} for the vibrational ground 
state (version 3 of entry 44504), which are mainly based on work by \citet{Mueller22}. 
Oxirane is well detected toward Sgr~B2(N2b), with two dozens of lines in its 
vibrational ground state easily identified (see Sec.\ 6 of the
\refsuppinfo). Integrated intensity maps of lines of 
oxirane that are free of contamination suggest an emission size on the order 
of \SI{0.7}{\arcsecond}. \autoref{f:popdiag_c2h4o-c_n2b} shows the population 
diagram of oxirane. A fit to this population diagram yields a rotational 
temperature of $132 \pm 10$~K. Assuming a temperature of 130~K, we adjusted a 
synthetic LTE spectrum to the observed spectrum and obtained the best-fit 
column density of oxirane reported in \autoref{t:coldens} that corresponds 
to the spectrum shown in red in Sec.\ 6 of the \refsuppinfo. 

\begin{table*}[!ht]
 \begin{center}
 \caption{
 Parameters of our best-fit LTE model of oxirane toward Sgr~B2(N2b) and upper limit for glycidaldehyde.
}
 \label{t:coldens}
 \resizebox{\linewidth}{!}{
 \begin{tabular}{lcrcccccccr}
 \hline\hline
 \multicolumn{1}{c}{Molecule} & \multicolumn{1}{c}{Status$^{(a)}$} & \multicolumn{1}{c}{$N_{\rm det}$$^{(b)}$} & \multicolumn{1}{c}{Size$^{(c)}$} & \multicolumn{1}{c}{$T_{\mathrm{rot}}$$^{(d)}$} & \multicolumn{1}{c}{$N$$^{(e)}$} & \multicolumn{1}{c}{$F_{\rm vib}$$^{(f)}$} & \multicolumn{1}{c}{$F_{\rm conf}$$^{(g)}$} & \multicolumn{1}{c}{$\Delta V$$^{(h)}$} & \multicolumn{1}{c}{$V_{\mathrm{off}}$$^{(i)}$} & \multicolumn{1}{c}{$\frac{N_{\rm ref}}{N}$$^{(j)}$} \\ 
  & & & \multicolumn{1}{c}{\small ($''$)} & \multicolumn{1}{c}{\small (K)} & \multicolumn{1}{c}{\small (cm$^{-2}$)} & & & \multicolumn{1}{c}{\small (km~s$^{-1}$)} & \multicolumn{1}{c}{\small (km~s$^{-1}$)} & \\ 
 \hline
 \ce{\textit{c}-C2H4O}, $\varv=0$$^\star$ & d & 28 &  0.7 &  130 &  4.0 (16) & 1.00 & 1.00 & 3.5 & 0.0 &       1 \\ 
 \ce{\textit{c}-(C2H3O)CHO}, $\varv=0$ & n & 0 &  0.7 &  130 & $<$  7.1 (15) & 1.43 & 1.00 & 3.5 & 0.0 & $>$     5.6 \\ 
\hline 
 \end{tabular}}
 \end{center}
\tnotes{
 $^{(a)}${d: detection, n: nondetection.}
 $^{(b)}${Number of detected lines \citep[conservative estimate, see Sect.~3 of][]{Belloche16}. One line of a given species may mean a group of transitions of that species that are blended together.}
 $^{(c)}${Source diameter (\textit{FWHM}).}
 $^{(d)}${Rotational temperature.}
 $^{(e)}${Total column density of the molecule. $x$ ($y$) means $x \times 10^y$.}
 $^{(f)}${Correction factor that was applied to the column density to account for the contribution of vibrationally excited states, in the cases where this contribution was not included in the partition function of the spectroscopic predictions.}
 $^{(g)}${Correction factor that was applied to the column density to account for the contribution of other conformers in the cases where this contribution could be estimated but was not included in the partition function of the spectroscopic predictions.}
 $^{(h)}${Linewidth (\textit{FWHM}).}
 $^{(i)}${Velocity offset with respect to the assumed systemic velocity of Sgr~B2(N2b), $V_{\mathrm{sys}} = 74.2$ km~s$^{-1}$.}
 $^{(j)}${Column density ratio, with $N_{\rm ref}$ the column density of the previous reference species marked with a $\star$.}
 }
 \end{table*}

\begin{figure*}
\centerline{\resizebox{0.85\hsize}{!}{\includegraphics[angle=0]{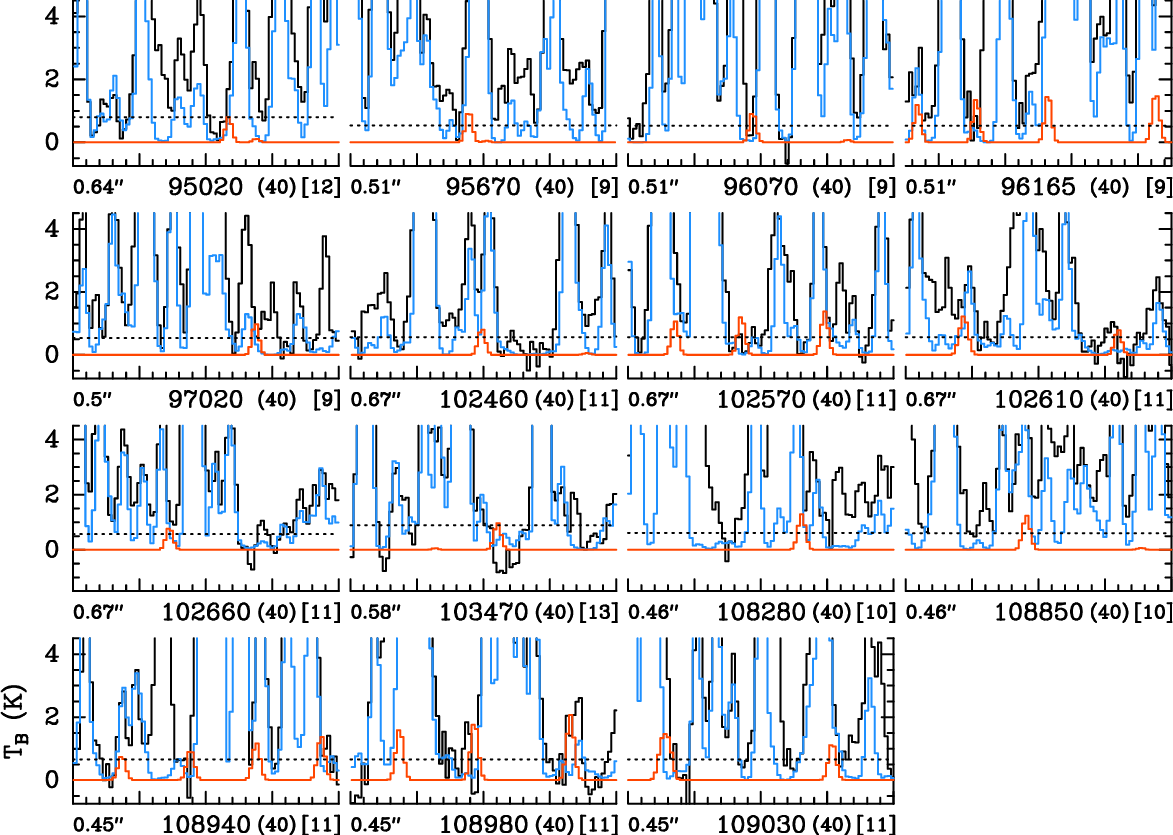}}}
\caption{Selection of rotational transitions of glycidaldehyde, 
\ce{\textit{c}-(C2H3O)CHO} covered by the ReMoCA survey. The LTE synthetic spectrum of 
\ce{\textit{c}-(C2H3O)CHO} used to derive the upper limit on its column density toward 
Sgr~B2(N2b) is displayed in red and overlaid on the observed spectrum shown in 
black. The blue synthetic spectrum contains the contributions of all molecules 
identified in our survey so far, but not the species shown in red. The 
values written below each panel correspond from left to right to the half-power 
beam width, the central frequency in MHz, the width in MHz of each panel in 
parentheses, and the continuum level in K of the baseline-subtracted spectra 
in brackets. The $y$-axis is labeled in brightness temperature units (K). The 
dotted line indicates the $3\sigma$ noise level.}
\label{f:remoca_glycidaldehyde_n2b}
\end{figure*}

To search for glycidaldehyde toward Sgr~B2(N2b) using the spectroscopic
predictions obtained in \autoref{sec:Glycidaldehyde}, we computed a synthetic LTE 
spectrum assuming the same velocity offset, linewidth, emission size, and 
rotational temperature as those derived for oxirane (\autoref{t:coldens}) 
and keeping the column density of glycidaldehyde as the only free parameter. 
We did not find any evidence for glycidaldehyde in this source. The synthetic 
spectrum used to estimate the upper limit to its column density is shown in 
red in \autoref{f:remoca_glycidaldehyde_n2b}. This upper limit is reported in 
\autoref{t:coldens}. We conclude from this analysis that glycidaldehyde is 
at least 6 times less abundant than oxirane in Sgr~B2(N2b).

\section{Conclusions}
\label{sec:Discussion}
In the present study, coverage of the experimental rotational spectrum of glycidaldehyde has been extended greatly by measuring \SI{345}{GHz} of broadband spectra in frequency intervals reaching as high as \SI{750}{GHz}.
The ground vibrational state, 17 vibrationally excited states, and the three singly substituted \ce{^{13}C} isotopologues were analyzed by combining the powers of LWPs and DM-DR measurements, which act as precise filters on the rather complicated spectrum.
Additionally, DM-DR measurements facilitated measuring interstate transitions between vibrationally excited states, yielding their vibrational energy separation with rotational precision.
In total, three interacting systems were examined and could be reproduced to about experimental accuracy.
The presented data allow for radio astronomical searches over a wide range of frequencies and quantum numbers.
Searches of glycidaldehyde's ground vibrational state with ALMA toward Sgr~B2(N2b) were not successful implying that glycidaldehyde is at least six times less abundant than oxirane in this source.
For future laboratory studies and astronomical searches of vibrationally excited states toward warm regions, the explicit interaction description in combined fits will be essential~\cite{Endres2021a}.
In the present study, foregoing the explicit interaction treatment and using only single-state fits would have resulted in about \SI{20}{\percent} of the approximately \SI{16000}{} total lines being rejected (due to $|\nu_\text{obs} - \nu_\text{calc}| / \Delta \nu > 10$) and considerably worse \textit{RMS} and \textit{WRMS} values.

Future studies might be targeted at filling the remaining frequency gaps and analyzing so-far untreated interactions.
Future steps for the DM-DR method include implementing a demodulation scheme with the sum (or difference) of the \textit{1f}- and \textit{2f}-demodulation frequencies or in the ideal case with a custom function being the product of a sine function with the \textit{2f}-demodulation frequency and a $\pm 1$ rectangle function for the subtraction.

Now that the rotational spectrum of glycidaldehyde is known very well, it might be worthwhile to check
for the status of other substituted oxiranes, \ce{\textit{c}-C2H3O-$X$}. A current census of the spectroscopic knowledge of six selected species has been given recently~\cite{Ellinger2020}. While rotational spectroscopic data have been collected for four of those ($X=$ \ce{CH3}, \ce{C2H}, \ce{CN}, and \ce{CHO}), at least two simple representatives (X=\ce{OH}, \ce{NH2}) still await their microwave spectroscopic characterization.

\section*{Data availability}
The input and output files of SPFIT will be provided as supplementary material.
These files as well as auxiliary files will be deposited in the \href{https://cdms.astro.uni-koeln.de/classic/predictions/daten/Glycidaldehyde/}{data section of the CDMS}.
Calculations of rotational spectra of the main isotopic species will be available in the \href{https://cdms.astro.uni-koeln.de/classic/entries/}{catalog section of the CDMS}.

\begin{suppinfo}
The calculated structures, rotation-vibration interaction constants, vibrational energies, and intensities.
Residuals histograms for assignments from this work and \citet{Creswell1977}
The transition shift due to interactions for one exemplary interaction system.
Parameter progression plot for the $v_{21}=n$ states.
Additional figure of the ReMoCA survey illustrating the detection of oxirane.
\end{suppinfo}

\begin{acknowledgement}
Glycidaldehyde studied in this work is an aldehyde derivative of glyceraldehyde, a sugar molecule, which according to the laboratory work of Prof.\ Dr.\ Harold Linnartz
is tentatively formed through recombination of reactive carbon bearing radicals along 
the well known \ce{CO} to \ce{CH3OH} hydrogenation route.
Therefore, this work is dedicated to the memory 
or our colleague and friend Harold Linnartz.
\newline
LB, ST, HSPM, and SS gratefully acknowledge the Collaborative Research Center 1601 (SFB 1601 sub-project A4) funded by the Deutsche Forschungsgemeinschaft (DFG, German Research Foundation) – 500700252.
JCG thanks the national program CNRS PCMI (Physics and Chemistry of the Interstellar Medium) and the CNES for a grant (CMISTEP).
This paper makes use of the following ALMA data: 
ADS/JAO.ALMA\#2016.1.00074.S. 
ALMA is a partnership of ESO (representing its member states), NSF (USA), and 
NINS (Japan), together with NRC (Canada), NSC and ASIAA (Taiwan), and KASI 
(Republic of Korea), in cooperation with the Republic of Chile. The Joint ALMA 
Observatory is operated by ESO, AUI/NRAO, and NAOJ. The interferometric data 
are available in the ALMA archive at https://almascience.eso.org/aq/.
\end{acknowledgement}

\bibliography{resources/bibliography,Arnaud/glycidaldehyde_remoca,resources/sthorwirth_bibdesk}

\end{document}